\documentclass[12pt]  {article}
\def\Journal#1#2#3#4{{#1}, {\bf #2} #3 (#4)}

\title{Classification of integrable polynomial vector evolution equations}
\author{Vladimir V.Sokolov \\
Landau Institute for Theoretical Physics, \\ 
Kosygina 2, Moscow 117334, Russia \\ 
email: sokolov@landau.ac.ru \\ 
\\
Thomas Wolf\\ 
Brock University, St.Catharines, \\
Ontario, Canada L2S 3A1 \\  
email: twolf@BrockU.ca}

\def\phi{\varphi}
\def\be{\begin{equation}}
\def\ee{\end{equation}}
\def\ba{\begin{array}}
\def\ea{\end{array}}
\def\bea{\begin{eqnarray}}
\def\eea{\end{eqnarray}}
\def\bean{\begin{eqnarray*}}
\def\eean{\end{eqnarray*}}

\newfam\msbfam
\font\tenmsb=msbm10
\textfont\msbfam=\tenmsb
\font\sevenmsb=msbm7
\scriptfont\msbfam=\sevenmsb
\font\fivemsb=msbm5 \scriptscriptfont\msbfam=\fivemsb

\def\Bbb#1{{\fam\msbfam\relax#1}}

\def\R{\Bbb R}  
\def\N{\Bbb N}  
  \def\Z{\Bbb Z}

\newfam\euffam
\font\teneuf=eufm10 \textfont\euffam=\teneuf
\font\seveneuf=eufm7 \scriptfont\euffam=\seveneuf

\begin{document}
\date{August 22, 2001}
\maketitle
\thispagestyle{empty}

\begin{abstract}
Several classes of systems of evolution equations with one or two vector unknowns are
considered. We investigate also systems with one vector and one scalar unknown. 
For these classes all equations having the simplest higher symmetry are listed.
\end{abstract}


\section{Introduction.}

The symmetry approach to the classification of integrable
evolution equations (see
\cite{SokSh,MikShYam,Fokas,MiShSok}) is proven to be the
most efficient integrability test for 1+1-dimensional
nonlinear PDEs.

In this paper we extend the simplest version of this
approach \cite{IbSh1,Fokas1} to the case of so called
vector evolution equations (see \cite{SvSok1}). As an
example let us consider the following vector equation
\be \label{mkdv1}
U_{t}=U_{xxx}+\langle U, U\rangle  U_x,
\ee
where $U(t,x)$ is a $N$-component vector and $\langle
\cdot,\cdot\rangle$ stands for the standard scalar product. If $N=1$
this formula reduces to the well-known modified Korteweg-de Vries (mKdV)
equation $u_t=u_{xxx}+u^2u_x$. Equation (\ref{mkdv1}) belongs to a wide
class \cite{SvSok1,Sv1,Sv2,Sv3,SvSok2,SokHabYam} 
of integrable multi-component evolution systems related to 
Jordan algebras and Jordan triple systems (see Appendix 1 for definitions).  

S.I. Svinolupov was the first who discovered a deep
relationship between some kinds of non-associative
algebraic structures and integrable multi-component
evolution systems. In the next two pages we formulate his results that have 
relations with the subject of our paper.

One of his results states that for any
Jordan triple system $\{X,Y,Z\}$  defined on a vector space $J$ the following 
equation
\be \label{jordmkdv}
U_{t}=U_{xxx}+\{ U, U,  U_x \},
\ee
where $U(x,t)$ is an unknown function with values in $J$ is integrable
(and, in particular, has higher symmetries). To derive some concrete
vector integrable systems from this general result we remind that there
exist two different "vector" Jordan triple systems:
\be \{X,Y,Z\}=\langle X,Y\rangle Z+
              \langle Y,Z\rangle X-
              \langle X,Z\rangle Y \label{jorvec2} \ee
and
\be  \{X,Y,Z\}=\langle X,Y\rangle Z+
               \langle Y,Z\rangle X. \label{jorvec1} \ee
Here $X,Y$ and $Z$ are arbitrary $N$-dimensional vectors. 
It is clear that (\ref{jorvec2}) gives rise (up to an inessential scaling
parameter) to (\ref{mkdv1})
while (\ref{jorvec1}) yields a different
integrable vector generalisation of the mKdV equation
\be \label{mkdv2}
U_{t}=U_{xxx}+ \langle U,  U\rangle \, U_x + \langle U,  U_x\rangle \, U.
\ee

The general formula (\ref{jordmkdv}) produces not only one-component
vector equations of mKdV-type. Let us consider the following Jordan triple
system 
\begin{equation}
\!\!
\begin{array}{l}
\left\{\left(\begin{array}{c}X^1\\X^2\end{array}\right),
       \left(\begin{array}{c}Y^1\\Y^2\end{array}\right),
       \left(\begin{array}{c}Z^1\\Z^2\end{array}\right)\right\}=\\[6mm]
\!\!
\left(\!\!\! \begin{array}{c}
(\langle Y^1,Z^1\rangle+\langle Y^2,Z^2\rangle )\,X^1+\!
(\langle X^2,Z^2\rangle-\langle X^1,Z^1\rangle )\,Y^1+\!
(\langle X^1,Y^1\rangle+\langle X^2,Y^2\rangle )\,Z^1 \\[3mm] 
(\langle Y^1,Z^1\rangle+\langle Y^2,Z^2\rangle )\,X^2+\!
(\langle X^1,Z^1\rangle-\langle X^2,Z^2\rangle )\,Y^2+\!
(\langle X^1,Y^1\rangle+\langle X^2,Y^2\rangle )\,Z^2
\end{array}\!\!\!\right)
\end{array} \label{twojord}
\end{equation}
where $X^1,Y^1,Z^1$ are vectors of length $N_1$ and $X^2,Y^2,Z^2$ are
vectors of length $N_2$. This Jordan triple system was not mentioned
in the original papers by Svinolupov. The corresponding system  
$$
\begin{array}{l}
U^1_{t}=U^1_{xxx}+\Big(\langle U^1,U^1 \rangle +
                         \langle U^2,U^2 \rangle\Big)\, U^1_{x}
                   +2 \langle U^2, U^2_{x}\rangle \, U^1\\[3mm]
U^2_{t}=U^2_{xxx}+\Big(\langle U^1,U^1 \rangle +
                         \langle U^2,U^2 \rangle \Big)\, U^2_{x}
                   +2 \langle U^1, U^1_{x}\rangle \, U^2 
\end{array}
$$   
with respect to vectors $U^1$ and $U^2$ of lengths $N_1$ and $N_2$
was presented in \cite{japan2}, although in non-explicit form this system
was probably found first in \cite{AtFor}. Notice that if $N_1=N_2=1$
then (\ref{twojord}) is a fully decoupled system in variables $\hat
U^1=U^1+U^2$\, and $ \hat U^2=U^1-U^2.$ For $N_1=N_2>1$ it is impossible to
split the system by this change of variables. 

One more general result by Svinolupov is related to a multi-component
generalisation of the nonlinear Schr\"{o}dinger equation (NLS) which can be
written as a system of two equations  
\be \label{NLS}
u_{t}=u_{xx}+ u^2 v, \quad v_t=-v_{xx}-v^2 u.
\ee
It was shown in \cite{Sv2} that for any Jordan triple system the
following system of equations 
\bea
\label{genls}
\cases{
U_t = \;\;\,U_{xx} + \{U,V,U\}, \vspace{6pt} \cr
V_t = - V_{xx} - \{V,U,V\} \cr}
\eea
is integrable. In particular, it has higher symmetries of any order  
$n\in \N$. For the vector triple systems (\ref{jorvec2}) and
(\ref{jorvec1}) the general formula (\ref{genls}) gives rise to two different 
integrable vector generalizations of (\ref{NLS}): 
\bea
\label{nls2}
\cases{
U_t = \;\;\,U_{xx} + 2 \langle U,V\rangle U  - 
       \langle U,U\rangle V, \vspace{6pt} \cr
V_t = - V_{xx} - 2 \langle U,V\rangle V + \langle V,V\rangle U, \cr}
\eea
and
\bea
\label{nls1}
\cases{
U_t = \;\;\,U_{xx} + \langle U,V\rangle U, \vspace{6pt} \cr
V_t = - V_{xx} - \langle U,V\rangle V. \cr}
\eea
Both systems are well known (see \cite{sclan,man}). 

Finally, Jordan triple system (\ref{twojord}) yields the 
following 4-vector system of NLS-type 
\bea
\cases{
U^1_{t} = \;\; U^1_{xx} + 2 (\langle U^1,V^1\rangle+
                          \langle U^2,V^2\rangle )\, U^1 
                     +   (-\langle U^1,U^1\rangle+
                          \langle U^2,U^2\rangle )\, V^1, \vspace{5pt} \cr
U^2_{t} = \;\; U^2_{xx} + 2 (\langle U^1,V^1\rangle+
                          \langle U^2,V^2\rangle )\, U^2 
                     +   (\langle U^1,U^1\rangle-
                          \langle U^2,U^2\rangle )\, V^2, \vspace{5pt} \cr
V^1_{t} = -V^1_{xx} - 2 (\langle U^1,V^1\rangle+
                           \langle U^2,V^2\rangle )\, V^1 
                      -   (-\langle V^1,V^1\rangle+
                           \langle V^2,V^2\rangle )\, U^1, \vspace{5pt} \cr
V^2_{t} = -V^2_{xx} - 2 (\langle U^1,V^1\rangle+
                           \langle U^2,V^2\rangle )\, V^2 
                      -   (\langle V^1,V^1\rangle-
                           \langle V^2,V^2\rangle )\, U^2, \cr},
\eea
where vectors $U^1$ and $V^1$ have length $N_1$ whereas $U^2$ and
$V^2$ are vectors of length $N_2$. It is easy to see that the
reduction $U^2=V^2=0$ reduces the system to (\ref{nls2}). If $N_1=N_2$
then the reduction $U^2=-U^1, \,V^2=-V^1$ becomes possible and results in system 
(\ref{nls1}). 

For all examples presented above, the right-hand side of the 
equation is a homogeneous differential polynomial under a
suitable weighting scheme. The differential equation
\be \label{equ}
u_t=f(u, u_x, \dots , u_{n-1}, u_n), \qquad u_i=\frac{\partial^i u}{\partial x^i}
\ee
is said to be $\lambda $-{\it homogeneous} of {\it weight} $\mu $ if it
admits the one-parameter group of scaling symmetries
$$(x, \ t, \ u)\longrightarrow (a^{-1}x, \ a^{-\mu} t, \ a^{\lambda} u).$$
For $N$-component systems with unknowns $u^1,...,u^N$ the corresponding
scaling group has a similar form
$$(x,t,u^1,...,u^N)\longrightarrow (a^{-1} x, \ a^{-\mu} t, \
a^{\lambda_1} u^1,...,
a^{\lambda_N} u^N).$$
In this paper we consider the case $\lambda_1=\cdots=\lambda_N$ only.

It was proven in \cite{SW} that in the scalar case a $\lambda$-homogeneous
polynomial equation with $\lambda>0$ may possess a homogeneous polynomial
higher symmetry only if
\begin{itemize}
\item Case 1: \qquad $\lambda=2$;
\item Case 2: \qquad $\lambda=1$;
\item Case 3: \qquad $\lambda={1\over 2}.$
\end{itemize}

For example, the KdV equation $u_t = u_{xxx} + u u_x$ is
homogeneous of weight $3$ for $\lambda =2$, the mKdV
equation $u_t = u_{xxx} + u^2 u_x $ has the weight $3$ for
$\lambda=1$ and for the Ibragimov-Shabat equation
\be \label{cal}
u_t = u_{xxx} + 3 u^2 u_{xx} + 9 u u_x^2 + 3 u^4 u_x
\ee
the weight is $3$ and $\lambda={1\over 2}$.

According to \cite{OlSok,OW} equations like (\ref{cal}) do
not exist in the matrix case. Thus the possible values of
$\lambda$ for matrix one-component equations possessing
higher symmetries are $2$ or $1$.

In the paper \cite{SW1} two-component systems of the form 
\bea \nonumber
\cases{
u_{t}=\;\;\, k_1 u_{xx}+P(u_x,v_x,u,v),\vspace{6pt} \cr
v_{t}=\;\;\, k_2 v_{xx}+Q(u_x,v_x,u,v) ,\cr}
\eea
where $k_1 \ne k_2, \, k_i \ne 0,$ 
have been considered. The authors 
assumed that $P$ and $Q$ are polynomials such that 
the system is $(\lambda_1,\lambda_2)$-homogeneous, 
where $\lambda_2\ge \lambda_1>0$ and 
$\lambda_2-\lambda_1$ is not a natural number. 
Under these assumptions it was proven that if 
the system possesses higher symmetries, then 
\begin{itemize}
\item if both $P$ and $Q$ have no quadratic terms, then $k_2=-k_1$;
\item if $\lambda_1=\lambda_2=\lambda$, then  
      $\lambda\in \{2,1,{1\over 2} \}.$
\end{itemize}
Also any possible pairs of $\lambda_1$ and $\lambda_2$ were listed.

For integrable systems with more than two unknowns
there are no rigorous statements describing all possible
types of homogeneous polynomial systems but for all known
examples with $\lambda_1=\cdots=\lambda_N=\lambda>0$,
$\lambda$ is equal to $2$, $1$ or ${1\over 2}$ just as in
the scalar and two-component cases.

In this paper we systematically investigate one and two component
homogeneous vector equations having higher symmetries under the assumption
$\lambda\in \{2,1,{1\over 2} \}$. It is remarkable that the 
number of different integrable cases for vector equations
turns out to be greater than in the matrix case.

Methodologically our approach is not new. But the continuous increase of 
computer speed and development of efficient computer algebra algorithms, 
especially for solving overdetermined algebraic systems allows to tackle problems 
which were unsolvable so far. 

Nevertheless, even modern technology cannot deal with
multicomponent systems, if we use a straightforward
component formalism. 
A corner stone of our approach in this paper 
is a special component-less representation of vectors
and vector equations.

Despite of such customised data-representations the
paper is the result of long and intensive computations
with the help of the computer algebra package {\sc Crack}
\cite{Wolf1, Wolf2}. Although originally designed to 
solve overdetermined PDE-systems, it got enhanced in recent years 
to solve algebraic and especially bi-linear systems efficiently
\cite{Wolf3}.

We find a number of new
vector models supposed to be integrable. The best way to
prove the integrability of these models is to find Lax
representations for them. We are planning to do that in a
separate paper.

\section{Integrable equations with one vector unknown}

In this section we consider vector equations of the form 
\begin{equation}
U_t=f_n \, U_n+f_{n-1}\, U_{n-1}+\cdots+f_1\,U_1+f_0\,U, 
    \qquad U_i=\frac{\partial^i U}{\partial x^i}
\label{gensys}
\end{equation}
with a single unknown vector-valued function $U$ of dimension $N$. Here 
$f_i$ are scalar functions of scalar products $\langle U_i,\, U_j\rangle, \, 
0\le i\le j \le n.$ We shall call equations of this form {\it isotropic}.  
It is clear that the isotropic equations are invariant
with respect to any orthogonal transformation of the vector $U$ i.e. the equation 
has the orthogonal group $O(N)$ as a group of point symmetries.

We shall consider equations (\ref{gensys}) that are
integrable for any dimension $N$. In addition, we
assume that the coefficients $f_i$ do not depend on $N$. In
virtue of the arbitrariness of $N,$ all different scalar
products $\langle U_i,\, U_j\rangle$ with $i\le j$ can be regarded as {\em
independent} variables. Their functional independence is a crucial
requirement in all our computations. If $N$ would be fixed, we
could not assume that. For instance, if $N=3,$ then the
determinant of the matrix with entries $a_{ij}=\langle U_i,\, U_j \rangle,
\, i,j=1,\ldots,4$  identically equals to zero.

The signature of the scalar product is unessential for us.
Furthermore, the assumptions that the vector space is
finite-dimensional and the constant field is $\R$ are
also unimportant. For instance, $U$ could be a function of
$t,x$ and $y$ and the scalar product be
$$
\langle U,\,V \rangle=\int^{\infty}_{-\infty} U(t,x,y)\, V(t,x,y)\, dy.
$$
In this way, our formulas and statements are valid also
for this particular sort of $1+2$-dimensional non-local
equations.

We suggest that the coefficients $f_i$ are {\em polynomials} in
$\langle U_i,\, U_j\rangle$ such that (\ref{gensys}) is homogeneous with
$\lambda=2$ (Case 1), $\lambda=1$ (Case 2) or $\lambda={1\over 2}$
(Case 3). Given the weight of (\ref{gensys}) it is easy to find the
most general form of such an equation (we have a special symbolic code
for that). Everywhere we assume that $f_n =\mbox{const} \ne 0$. 

For example, the general form for an equation of weight 3 with $\lambda=1$ is
given by
\be \label{genmkdv}
U_t=a_1 U_{xxx}+a_2 \langle U,U\rangle U_x + a_3 \langle U, U_x\rangle U,
\ee
where $a_i$ are arbitrary constants, $a_1\ne 0.$ Using a scaling of $t$ we
can normalize $a_1$ to $1$.

To describe all integrable cases we assume (cf. with the scalar case
\cite{IbSh1,Fokas,OlSok,SW}) that (\ref{genmkdv}) has a higher homogeneous
symmetry of weight 5, whose general form is
\begin{eqnarray}
 U_{\tau} &=& U_{xxxxx} + b_1 \langle U,U  \rangle \ U_{xxx} 
                        + b_2 \langle U,U_x\rangle \ U_{xx} + \nonumber \\
          & &  (b_3 \langle U,U_{xx}\rangle+
                b_4 \langle U_x,U_x \rangle+
                b_5 \langle U,U\rangle^2    ) \ U_x + \label{gensymmkdv} \\
          & &  (b_6 \langle U, U_{xxx}\rangle+
                b_7 \langle U_x, U_{xx}\rangle+
                b_8 \langle U,U\rangle\langle U,U_x\rangle) \ U. \nonumber
\end{eqnarray}

Compatibility conditions $(U_{t})_{\tau}=(U_{\tau})_t$ of 
(\ref{genmkdv}) and (\ref{gensymmkdv}) give rise to a system of 
26 bilinear algebraic equations with respect to unknowns $a_2, a_3,
b_1,\dots,b_8$. Solving this system with the help of the
symbolic computer program {\sc Crack} we finally obtain
that the system has a solution only if $a_3=0$ or $a_2=-2
a_3$. These two possibilities lead to integrable equations
(\ref{mkdv1}) and (\ref{mkdv2}).

We considered Cases 1,2 and 3 corresponding to different weighting
schemes. For each case we have performed the similar calculations to
find all equations of second order having a symmetry of third order,
equations of third order having a symmetry of fifth order, and
equations of fifth order possessing a symmetry of seventh order. The complexity 
of the algebraic system increases drastically with an increase of the total weight 
(differential order of the equation and symmetry) and with decreasing $\lambda$. 
For details see Appendix 2.

It turns out that apart from equations (\ref{mkdv1}) and (\ref{mkdv2}),
there exists only one more integrable equation:
\be
\label{ibsh}
U_{t}=U_{xxx}+3 \langle U, U\rangle U_{xx}+6 \langle U,U_x\rangle U_x+
              3 \langle U,U\rangle^2 U_x+3 \langle U_x, U_x\rangle U
\ee
with $\lambda=\frac{1}{2}$. This vector analogue of the Ibragimov-Shabat
equation (\ref{cal}) was already found in \cite{SokWol}.

All equations (\ref{mkdv1}),(\ref{mkdv2}), and (\ref{ibsh}) belong to
infinite commutative hierarchies of evolution equations such that any
equation from the hierarchy is a symmetry for all others. These three
hierarchies are of the same (3,5)-type. This notation means that the  
simplest non-trivial equation in the hierarchy is of order three and
the next is an equation of order 5.

We see, that there are no hierarchies of type (2,3) and
(5,7) in the vector case with one unknown vector in
contrast with the scalar case, where such hierarchies
exist (Burgers, Kaup-Kupershmidt, Kupershmidt and
Sawada-Kotera equations). 

Combining our results with the approach by Sanders-Wang \cite{SW,OW}  
one can prove that any polynomial homogeneous equation of the form
(\ref{gensys}) with $\lambda>0$ and $f_n =const \ne 0$ which has at 
least one higher symmetry, belongs to one of the above three hierarchies
\cite{san}. 

\section{Integrable equations with two vector unknowns}

In this section we classify integrable vector NLS-type
systems. More precisely, we consider systems
of the form
\bea
\label{nceqgen}
\cases{
U_{t}=\;\;\,U_{xx}+p_1 U_x+p_2 V_x+p_3 U+p_4 V,\vspace{6pt} \cr
V_{t}=-V_{xx}+p_5 U_x+p_6 V_x+p_7 U+p_8 V ,\cr}
\eea
where $U$ and $V$ are vectors and the coefficients $p_i$ are
$\lambda$-homogeneous polynomials depending on all possible scalar products of vectors
$U,V,U_x,V_x$. By analogy with the scalar case $\lambda$ is supposed to be 2,1, or 
$\frac{1}{2}$. To derive the integrable cases 
we assume that (\ref{nceqgen}), just as in the scalar case (see \cite{MikShYam}) 
possesses a symmetry of the form
\bea
\label{symgen}
\cases{
U_{\tau}=U_{xxx}+q_1 U_{xx}+q_2 V_{xx}+q_3 U_x+q_4 V_x+q_5 U+q_6 V,\vspace{6pt} \cr
V_{\tau}=V_{xxx}+q_7 U_{xx}+q_8 V_{xx}+q_9 U_x+q_{10} V_x+q_{11} U+q_{12} V,
\cr}
\eea
where the coefficients $q_i$ are $\lambda$-homogeneous 
polynomials of all possible scalar products of $U,V,U_x,V_x, U_{xx}, V_{xx}$.
We call such systems integrable. The classification result is the following.

It is easy to see that nonlinear systems (\ref{nceqgen}) corresponding to the case $\lambda=2$ do not exist. 

For $\lambda=1$ there exist only two 
integrable systems (\ref{nls1}) and (\ref{nls2}) (up to a scaling of $U,V,x$ and $t$).

In the case $\lambda=\frac{1}{2}$ the complete list of integrable cases
(up to the scaling and the involution
$U \leftrightarrow V, \ t \leftrightarrow -t$)
looks as follows:
\bea
\label{meq1}
\cases{
U_t = \;\;\,U_{xx} + 2 \alpha \langle U, V\rangle U_x 
                   + 2 \alpha \langle U, V_x\rangle U
                   - \alpha \beta \langle U,V\rangle^2 U, \vspace{5pt} \cr
V_t =   -   V_{xx} + 2 \beta \langle U, V\rangle V_x 
                   + 2 \beta \langle V, U_x\rangle V
                   + \alpha \beta \langle U,V\rangle^2 V  \cr}
\eea

\bea
\label{meq2}
\cases{
U_t = \;\;\,U_{xx} + 2 \alpha \langle U, V\rangle U_x 
                   + 2 \beta \langle U, V_x\rangle U
                   + \beta (\alpha -2 \beta) \langle U,V\rangle^2 U, \vspace{5pt}\cr
V_t =   -   V_{xx} + 2 \alpha \langle U, V\rangle V_x 
                   + 2 \beta \langle V, U_x\rangle V
                   - \beta (\alpha -2 \beta) \langle U,V\rangle^2 V  \cr}
\eea
\bea
\label{meq3}
\cases{
U_t = \;\;\,U_{xx} + 2 \alpha \langle U, V\rangle U_x 
                   + 2 \beta \langle U, V_x\rangle U 
                   + 2 (\beta-\alpha) \langle V, U_x\rangle U  \cr
\;\;\;\;\;\;\;\;- \alpha \beta \langle U,V\rangle^2 U, \vspace{5pt} \cr
V_t =   -   V_{xx} + 2 \alpha \langle U, V\rangle V_x 
                   + 2 \alpha \langle V, U_x\rangle V
                   + \alpha \beta\langle U,V\rangle^2 V \cr}
\eea

\bea
\label{meq4}
\cases{
U_t = \;\;\,U_{xx} + 2 \alpha \langle U, V\rangle U_x 
                   + 2 \alpha \langle U, V_x\rangle U 
                   + 2 \beta \langle V, U_x\rangle U \cr
\;\;\;\;\;\;\;\;-  \alpha (\alpha-\beta) \langle U,V\rangle^2 U, \vspace{5pt} \cr
V_t =   -   V_{xx} + 2 \alpha \langle U, V\rangle V_x 
                   + 2 \alpha \langle V, U_x\rangle V
                   + 2 \beta \langle U, V_x\rangle V \cr
\;\;\;\;\;\;\;\;+ \alpha (\alpha-\beta) \langle U,V\rangle^2 V \cr}
\eea

\bea
\label{meq5}
\cases{
U_t = \;\;\,U_{xx} + 4 \alpha \langle U, V\rangle U_x 
                   + 2 (\alpha-\beta) \langle U, U\rangle V_x 
                   + 4 \beta \langle U, V_x\rangle U \cr
\;\;\;\;\;\;\;\;   + 4 \beta (\alpha- 2\beta) \langle U,V\rangle^2 U
                   - 2 \beta (\alpha-\beta) \langle U,U\rangle
                                            \langle V,V\rangle U \cr
\;\;\;\;\;\;\;\;   - 4 \beta (\alpha-\beta) \langle U,U\rangle
                                            \langle U,V\rangle V, \vspace{5pt} \cr
V_t =   -   V_{xx} + 4 \alpha \langle U, V\rangle V_x 
                   + 2 (\alpha-\beta) \langle V, V\rangle U_x
                   + 4 \beta \langle V, U_x\rangle V \cr
\;\;\;\;\;\;\;\;   - 4 \beta (\alpha- 2\beta) \langle U,V\rangle^2 V 
                   + 2 \beta (\alpha-\beta) \langle U,U\rangle
                                            \langle V,V\rangle V \cr
\;\;\;\;\;\;\;\;   + 4 \beta (\alpha-\beta) \langle V,V\rangle
                                            \langle U,V\rangle U \cr}
\eea

\bea
\label{meq6}
\cases{
U_t = \;\;\,U_{xx} + 4 \alpha \langle U, V\rangle U_x 
                   - 2 \beta  \langle U, U\rangle V_x
                   + 4 \alpha \langle U,V_x\rangle U \cr 
\;\;\;\;\;\;\;\;   + 4 \beta \langle V, U_x\rangle U 
                   - 4 \beta \langle U, U_x\rangle V 
                   + 6 \beta (\alpha - \beta) \langle U,U\rangle
                                              \langle V,V\rangle U \cr
\;\;\;\;\;\;\;\;   - 4 \alpha (\alpha - \beta) \langle U,V\rangle^2 U
                   - 4 \beta  (\alpha - \beta) \langle U,U\rangle
                                               \langle U,V\rangle V, \vspace{5pt} \cr
V_t =   -   V_{xx} + 4 \alpha \langle U, V\rangle V_x 
                   - 2 \beta \langle V, V\rangle U_x
                   + 4 \alpha \langle V,U_x\rangle V \cr 
\;\;\;\;\;\;\;\;   + 4 \beta \langle U, V_x\rangle V 
                   - 4 \beta \langle V,V_x\rangle U 
                   - 6 \beta (\alpha- \beta) \langle U,U\rangle
                                             \langle V,V\rangle V \cr
\;\;\;\;\;\;\;\;   + 4 \alpha (\alpha - \beta) \langle U,V\rangle^2 V 
                   + 4 \beta  (\alpha - \beta) \langle V,V\rangle
                                               \langle U,V\rangle U \cr}
\eea

All equations in the list contain arbitrary constants
$\alpha$ and $\beta$. It is easy to see
that if $\alpha$ is not equal to zero, then it
can be reduced to 1 via scalings of $t,x,U$ and $V$.
Thus the essential parameter is the ratio of
$\beta$ and $\alpha$. We choose the above form with the two arbitrary
constants to avoid considering the case $\alpha=0$ separately.

{\bf Remark 1.} In the paper \cite{japan1} equations (\ref{meq2}) and (\ref{meq4}) 
were found and Lax representations for them were presented. The other equations 
seem to be new. 

{\bf Remark 2.} In the paper \cite{olvsok} a list of matrix 
integrable equations have been obtained. This list contains all integrable 
polynomial homogeneous matrix equations with $\lambda=\frac{1}{2}$. 
Some particular cases of our vector systems (\ref{meq1})-(\ref{meq6}) 
can be derived from the matrix list as reductions. 
For example, for all matrix systems we can perform the following reduction 
$$
U=\left( \begin{array}{ccccc}
0&0&...&0&u^1\\
0&0&...&0&u^2\\
...&...&...&...&...\\
0&0&...&0&u^N\\
v^1&v^2&...&v^N&0
\end{array}
\right), \qquad 
V=\left( \begin{array}{ccccc}
0&0&...&0&x^1\\
0&0&...&0&x^2\\
...&...&...&...&...\\
0&0&...&0&x^N\\
y^1&y^2&...&y^N&0
\end{array}
\right)
$$
to get the corresponding systems with four vector unknowns $U,V,X,Y$. Sometimes 
we can decrease the number of vector unknowns to two by further reductions and obtain some 
systems from the list (\ref{meq1})-(\ref{meq6}). But since non of the matrix equations from 
\cite{olvsok} have arbitrary parameters and our equations do, we probably can construct 
only particular equations from the list (\ref{meq1})-(\ref{meq6}) 
by reductions of the matrix equations.

\section{Integrable equations with one scalar and one vector unknown}

Some vector analogues of the KdV equation can be constructed using a different
general result of Svinolupov \cite{Sv1}: for any Jordan 
algebra (see the definition in the Appendix 1) with
multiplication $\circ$ the following "Jordan" KdV equation
\be \label{jordkdv}
U_{t}=U_{xxx}+ U \circ U_x
\ee
has higher symmetries. Since for any Jordan triple system $J=\{X,Y,Z\}$ and
given fixed vector $C$ the following multiplication
$$
X  \circ  Y=\{X,  C,  Y \}
$$
defines a Jordan algebra, the formulas (\ref{jorvec2}) and (\ref{jorvec1})
produce two vector KdV equations
\be \label{kdv2}
U_{t}=U_{xxx} + \langle U,  C\rangle \ U_x 
              + \langle C,  U_x\rangle \ U 
              - \langle U,  U_x\rangle \ C
\ee
and
\be \label{kdv1}
U_{t}=U_{xxx}+ \langle U, C\rangle \, U_x + \langle C, U_x\rangle \, U.
\ee
The first had been obtained in \cite{SvSok1} and the second
was considered in \cite{AtFor}.

The Jordan triple system (\ref{twojord}) gives rise to 

\begin{equation} \left\{
\begin{array}{ll}
U^1_{t} = U^1_{xxx} + & ( \langle U^1,C^1 \rangle+
                            \langle U^2,C^2 \rangle )\, U^1_{x}+ \vspace{3pt}\\
                        & ( \langle C^1,U^1_{x} \rangle+
                            \langle C^2,U^2_{x} \rangle )\, U^1 + \vspace{3pt}\\
                        & (-\langle U^1,U^1_{x} \rangle+
                            \langle U^2,U^2_{x} \rangle )\, C^1, \vspace{6pt}\\
U^2_{t} = U^2_{xxx} + & ( \langle U^1,C^1 \rangle+
                            \langle U^2,C^2 \rangle)\, U^2_{x}+  \vspace{3pt}\\
                        & ( \langle C^1,U^1_{x} \rangle+
                            \langle C^2,U^2_{x} \rangle )\, U^2 + \vspace{3pt} \\
                        & ( \langle U^1,U^1_{x}\rangle-
                            \langle U^2,U^2_{x}\rangle )\, C^2, 
\end{array}      \right.
\label{doublenls}
\end{equation}

Note that it is impossible to construct any vector KdV equation
(with a term quadratic in $U$) without
using constant vectors. Indeed, the simplest 
nonlinear vector term which
can be generated with the help of the scalar product
from only $U, U_x, U_{xx}$ is of third degree.

For non-isotropic equations like (\ref{kdv1}) one can use the orthogonal
group $O(N)$ to simplify the vector $C$ and bring it, for example, to the form
\be \label{CC}
C=(1,0,0,...,0).
\ee
With vector $C$ fixed, the equation admits the symmetry
group $O(N-1)$.

The component form of (\ref{kdv1}) with $C$ given by (\ref{CC})
is the following
\bea
\nonumber
\cases{
u^1_t = u^1_{xxx} + 2  u^1 u^1_x, \vspace{4pt}\cr
u^2_t = u^2_{xxx} + u^1 u^2_x + u^1_x u^2, \vspace{4pt} \cr
...\cr
u^N_t = u^N_{xxx} + u^1 u^N_x + u^1_x u^N, \cr }
\eea
where $U=(u^1,\dots,u^N).$ We see that the first component
$u^1$ satisfies the standard scalar KdV equation and the
remaining equations are linear if the function $u^1$ is
already found. We call such vector systems {\it
triangular}.

In contrast, (\ref{kdv2}) cannot be splited by any
orthogonal transformation, i.e. it is {\it
non-triangular}. Note that if $C$ is of the form
(\ref{CC}) then (\ref{kdv2}) can be rewritten as
\bea
\label{scveckdv}
\cases{
u_t = u_{xxx} + u u_x - \langle U, U_x\rangle, \vspace{5pt} \cr
U_t = U_{xxx} + u U_x + u_x U, \cr}
\eea
where $u=u^1$ and $U=(u^2,\dots,u^N).$

System (\ref{scveckdv}) gives us an example of an 
integrable systems with one vector and one scalar
unknown. Surprisingly, for such systems the list of
integrable models seems to be much richer than in the 
case of isotropic equations. In the following
a lower case $u$ stands for the scalar variable
and the vector variable is denoted by a capital $U$.

\subsection{The case $\lambda=2$}

In this paper we consider the simplest case $\lambda=2$. The
generic form for a second order homogeneous system of this
type reads as follows
\begin{equation}
\label{gen2-2}
\cases{
\begin{array}{rcl}
u_t&=&a_1 \, u_{xx} + a_2\, u^2+ a_3\,\langle U, \, U\rangle, \\[3mm]
U_t&=& a_4 \, U_{xx} + a_5 \, u\,U,
\end{array}
}
\end{equation}
where $a_i$ are constants.  We assume that at least one of
the constants $a_1$ or $a_4$ is non-zero. It means that
the system is non-degenerate. As in the scalar case, the
classification result is negative. It turns out that
system (\ref{gen2-2}) has a symmetry of third order only
if $a_2=a_3=a_5=0.$

For third order systems with $\lambda=2$ the result is
much more interesting. The generic form of such systems is
given by
\begin{equation}
\label{gen3-2}
\cases{
\begin{array}{rcl}  
u_t&=&a_1 \, u_{xxx} + a_2\, u u_x+ a_3\,\langle U, \, U_x\rangle,
\\[3mm] 
U_t&=& a_4 \, U_{xxx} + a_5 \, u\,U_x+a_6 \, u_x \, U.
\end{array}
}
\end{equation}
These systems can be regarded as natural vector
generalizations of the KdV equation. We assume that \, 
1) the system is of third order, i.e.\ 
at least one of the constants $a_1$ or $a_4$ is non-zero; \, 
2) our system is non-triangular, i.e.\ 
$a_3\ne 0$ and one of $a_5$ or $a_6$ is non-zero. 

{\bf Theorem.} Suppose a non-triangular system
(\ref{gen3-2}) possesses a symmetry of fifth order. Then
(up to a scaling of $t,x,u$ and $U$) this system 
coincides with (\ref{scveckdv}), or belongs to the
following list

\begin{equation}
\label{eeq2V}
\cases{
\begin{array}{rcl}
u_t&=&u_{xxx} + 3 u u_x + 3 \langle U, U_x\rangle, \\[3mm]
U_t&=&u \, U_x + u_x \, U,
\end{array}
}
\end{equation}

\begin{equation}
\label{eeq3V}
\cases{
\begin{array}{rcl}
u_t&=&\langle U, \,  U_x\rangle, \\[3mm] 
U_t&=&U_{xxx} + 2 u \, U_x + u_x \, U,
\end{array}
}
\end{equation}

\begin{equation}
\label{eeq4V}
\cases{
\begin{array}{rcl}
u_t&=&u_{xxx} + u u_x + \langle U, U_x\rangle, \\[3mm]
U_t&=&-2\,U_{xxx} - u \, U_x.
\end{array}
}
\end{equation}

System (\ref{eeq2V}) has been considered by Kupershmidt 
\cite{kuper}.  It is a vector generalisation of the Ito
equation.

Systems (\ref{eeq3V}) and (\ref{eeq4V}) seem to be new.
They are vector generalisations of the corresponding
scalar systems from the paper \cite{DriSok}. A coherent
picture of such vector analogues of the Kac-Moody
KdV-systems will be developed in a separate paper. It is
interesting to note that the fifth order symmetry
\begin{equation}
\label{kuper}
\begin{array}{rcl}
u_t&=&u_{xxxxx} + 10 u u_{xxx}+25 u_x u_{xx}+20 u^2 u_x-\\[3mm]
   & &10 \langle U, U_{xxx}\rangle- 15 \langle U_x, U_{xx}\rangle  
      -10 u_x \langle U,U\rangle-20 u \langle U, U_x\rangle,\\[6mm] 
U_t&=&-9\,U_{xxxxx} - 30 u \, U_{xxx}-45 u_x \, U_{xx}
      -(35 u_{xx}+20 u^2+5 \langle U,U\rangle)\,U_x\\[3mm]
   & &-(10 u_{xxx}+20 u u_x+5 \langle U,U_x\rangle)\,U
\end{array}
\end{equation}
of system (\ref{eeq3V}) is nothing but a vector
generalization of the Kaup-Kuperschmidt equation. Indeed,
if the vector part is absent (i.e. $U=0$) then system
(\ref{eeq3V}) becomes trivial and (\ref{kuper}) turns out
to be the Kaup-Kuperschmidt equation.

Our attempts to find vector analogues for the
Sawada-Kotera equation among systems with one vector and
one scalar unknowns were unsuccessful. It turns out
that any non-triangular, non-degenerate system of fifth
order with $\lambda=2$, having a symmetry of seventh
order, is a symmetry of one of (\ref{scveckdv}),
(\ref{eeq2V}), (\ref{eeq3V}) or (\ref{eeq4V}).

\section*{Appendix 1.}
{\bf Definition of Jordan algebra:} A vector space $J$ 
equipped with a product $\circ: J\times J\rightarrow J$ such that 
$$
X\circ Y=Y\circ X, \qquad X^2\circ (Y\circ X)=(X^2\circ Y)\circ X
$$
for any $X,Y\in J$ is called {\it Jordan algebra}.

If $*$ is a multiplication in an associative algebra then $X\circ Y=X*Y+Y*X$ 
is a Jordan product. \newline
{\bf Definition of Jordan triple system:} A vector 
space $J$ equipped with a triple product 
$\{\cdot,\cdot,\cdot\}: J\times J\times J \rightarrow J$
such that 
$$
\{X,Y,Z\}=\{Z,Y,X\}
$$
$$
\{X,Y,\{V,W,Z\}\}-\{V,W,\{X,Y,Z\}\}=
\{\{X,Y,V\},W,Z\}-\{V,\{Y,X,W\},Z\}
$$
for any $V,W,X,Y,Z\in J$ is called {\it Jordan triple system}.

Any associative algebra with a multiplication $*$ produces  
a Jordan triple system with respect to the triple product  
$$
\{X,Y,Z\}=X*Y*Z+Z*Y*X.
$$

\section*{Appendix 2.}
For different ans\"{a}tze of a vector equation and its symmetry, like 
equations (\ref{genmkdv}) and (\ref{gensymmkdv}), the algebraic conditions
for the unknown coefficients vary drastically in their size. For the 
nine combinations of 3 values $\lambda=2,1,{1 \over 2}$ and 
3 pairs of differential order of the equation and its symmetry 
((2+3),(3+5),(5+7)) the table below shows the number of unknowns,
the number of conditions and the total number of terms in these
conditions for the resulting nine over-determined systems of
algebraic conditions. 

\begin{center}
\begin{tabular}{|c||c||c||c||r|} \hline
 order $\rightarrow$
               & 2 + 3 & 3 + 5 & 5 + 7 &                   \\ \hline\hline
               &   0   &   2   &   8   & \# of unknowns     \\ \cline{2-5}
 $\lambda=2$   &   0   &   4   &   8   & \# of equations    \\ \cline{2-5}
               &   0   &   4   &   8   & total \# of terms  \\ \hline\hline
               &   3   &  10   &   31  & \# of unknowns     \\ \cline{2-5}
 $\lambda=1$   &   5   &  26   &  198  & \# of equations    \\ \cline{2-5}
               &   9   &  121  & 3125  & total \# of terms  \\ \hline\hline
               &  10   &  33   &  107  & \# of unknowns     \\ \cline{2-5}
 $\lambda={1 \over 2} $ 
               &  21   &  129  &  927  & \# of equations    \\ \cline{2-5}
               &  80   & 1603  & 52677 & total \# of terms  \\ \hline
\end{tabular} \vspace{6pt}\\
Table 1. A comparison of the size of the symmetry conditions for single
vector equations.
\end{center}

\section*{Acknowledgement}
The authors are grateful to Jan Sanders and Jing Ping Wang for 
useful discussions. They are also thankful to Takayuki Tsuchida
for comments about the integrability of systems found.
The first author (V.S.) was supported, in part, by RFBR grant
99-01-00294, INTAS grant 99-1782, and EPSRC grant GR K99015.
The second author (T.W.) thanks Winfried Neun for 
computer algebra related discussions and
the UMS MEDICIS service at the CNRS/Polytechniqe in computer algebra
for providing access to their computing facilities. 
Both authors thank the Newton Institute, Cambridge for its hospitality 
during the programme "What is Integrability" where the manuscript 
was completed. 


\end{document}